\documentclass[prd,onecolumn,nofootinbib]{revtex4}
\newcommand\gothfamily{\usefont{U}{ygoth}{m}{n}}
\DeclareTextFontCommand{\textgoth}{\gothfamily}

\begin{document}

\title{MASSIVE VECTORS FROM PROJECTIVE-INVARIANCE BREAKING}

\author{{\bf Nikodem J. Pop\l awski}}

\affiliation{Department of Physics, Indiana University, Swain Hall West, 727 East Third Street, Bloomington, Indiana 47405, USA}
\email{nipoplaw@indiana.edu}

\noindent
{\em Annales de la Fondation Louis de Broglie}\\
Vol. {\bf 32}, No. 2-3 (2007) 335--353\\
\vspace{0.1in}

\begin{abstract}
ABSTRACT.
A general affine connection has enough degrees of freedom to describe the classical gravitational and electromagnetic fields in the metric--affine formulation of gravity.
The gravitational field is represented in the Lagrangian by the symmetric part of the Ricci tensor and the classical electromagnetic field can be represented by the tensor of homothetic curvature.
The simplest metric--affine Lagrangian that depends on the tensor of homothetic curvature generates the Einstein--Maxwell equations for a massless vector.
Metric--affine Lagrangians with matter fields depending on the connection are subject to an unphysical constraint because the symmetrized Ricci tensor is projectively invariant while matter fields are not.
We show that the appearance of the tensor of homothetic curvature, which is not projectively invariant, in the Lagrangian replaces this constraint with the Maxwell equations and restores projective invariance of the total action.
We also examine several constraints on the torsion tensor to show that algebraic constraints on the torsion that break projective invariance of the connection and impose projective invariance on the tensor of homothetic curvature replace the massless vector with a massive vector.
We conclude that the metric--affine formulation of gravity allows for a mechanism that generates masses of vectors, as it happens for electroweak gauge bosons via spontaneous symmetry breaking.
\end{abstract}

\maketitle

\section{Introduction}
\label{secIntro}

There exist three formulations of general relativity: {\em purely metric}, {\em metric--affine} and {\em purely affine}~\cite{FK3}.
In the purely metric (Einstein--Hilbert) formulation, the metric tensor is a variable (gravitational potential) and the field equations are derived by varying the total action with respect to the metric.
The affine connection is the Christoffel--Levi-Civita connection associated with the metric and the Lagrangian density for the gravitational field is linear in the symmetric part of the Ricci tensor.
In the metric--affine (Einstein--Palatini) formulation, both the metric tensor and a torsionless connection are independent variables (gravitational potentials) and the field equations are derived by varying the action with respect to these quantities.
The Lagrangian for the gravitational field is linear in the symmetric part of the Ricci tensor.
In the purely affine (Einstein--Eddington) formulation, a Lagrangian density depends on a torsionless affine connection and the symmetric part of the Ricci tensor.
This formulation constructs the (symmetric) metric tensor as the Hamilton derivative of the Lagrangian density with respect to the symmetrized Ricci tensor, obtaining an algebraic relation between these two tensors.
The field equations are derived by varying the action with respect to the connection, which gives a differential relation between the connection and the metric tensor.
This relation yields a differential equation for the metric.

Ferraris and Kijowski showed that all three formulations of general relativity are dynamically equivalent and the relation between them is analogous to the Legendre relation between Lagrangian and Hamiltonian dynamics~\cite{FK3}.
The relation between the purely affine and metric--affine picture of general relativity shows that the metric--affine Lagrangian density for the gravitational field is a Legendre term (with respect to the symmetrized Ricci tensor) corresponding to the scalar product of the velocities and momenta in classical mechanics~\cite{Kij}.
The purely metric and metric--affine picture of general relativity are related by a Legendre transformation with respect to the connection, and the Lagrangian density for the gravitational field in these pictures automatically turns out to be linear in the curvature tensor~\cite{FK3}.
The equivalence of the three formulations can also be shown for theories of gravitation with purely affine Lagrangians that depend on the full Ricci tensor and the tensor of homothetic curvature~\cite{univ,nonsym}, and a general connection with torsion~\cite{nonsym,Smal}.
There also exist formulations of gravity in which the dynamical variables are: metric and torsion (Einstein--Cartan theory)~\cite{Hehl,SG}, metric and nonsymmetric connection~\cite{HK,HLS}, tetrad~\cite{Uti}, tetrad and spin connection (Einstein--Cartan--Kibble--Sciama theory)~\cite{KS}, pure spin connection~\cite{sc}, and spinors~\cite{sp}.

Both the purely affine and metric--affine picture yield the same field equation from varying the action with respect to the connection.
The difference between these two formulations is the role of the metric.
In the purely affine picture the metric is defined from the Lagrangian density as a gravitational momentum, while in the metric--affine picture the metric is a variable (configuration)~\cite{FK3}.
Accordingly, the algebraic relation between the Ricci tensor and the metric tensor follows from a Hamilton-derivative definition in the purely affine picture and from a variational principle in the metric--affine picture.
In this paper we examine only the field equations obtained from varying the action with respect to the connection.
Therefore our results are valid for both the purely affine and metric--affine picture.

In general relativity, the electromagnetic field and its sources are considered to be on the side of the matter tensor in the field equations, i.e. they act as sources of the gravitational field.
In unified field theory, the electromagnetic field obtains the same geometric status as the gravitational field~\cite{Goe}.
A general affine connection, not restricted to be metric-compatible and symmetric, has enough degrees of freedom to make it possible to describe the classical gravitational and electromagnetic fields.
Both the purely affine and metric--affine formulation of gravity allow an elegant unification of these fields.
Ferraris and Kijowski showed that while the gravitational field in the purely affine formulation is represented by the symmetric part of the Ricci tensor, the electromagnetic field can be represented by the tensor of homothetic curvature~\cite{FK2}.
The purely affine Lagrangian for the electromagnetic field, that has the form of the Maxwell Lagrangian in which the metric tensor is replaced by the symmetrized Ricci tensor and the electromagnetic field tensor by the tensor of homothetic curvature, is dynamically equivalent to the sourceless Einstein--Maxwell Lagrangian in the metric--affine and metric formulation~\cite{FK2,FK1}.
Such a construction can be generalized to sources~\cite{unif}.
Ponomarev and Obukhov introduced the same construction in the metric--affine formulation of gravity~\cite{PO}.

Affine (metric--affine and purely affine) Lagrangians that depend on the affine connection not only through the symmetric part of the Ricci tensor but also through matter fields, are subject to an unphysical constraint on the source density~\cite{HK,HLS,San}.
This constraint is related to the fact that the symmetrized Ricci tensor is projectively invariant~\cite{HK,HLS,San,Scho} while the matter part of the Lagrangian is generally not.
The inclusion of the tensor of homothetic curvature (the electromagnetic field), which is also projectively non-invariant, in an affine Lagrangian replaces this constraint with the Maxwell equations and restores projective invariance of the total action without constraining the connection~\cite{unif}.

In this paper we consider the {\em metric--affine} formulation of gravity.
In Sec.~\ref{secField} we show, following Ponomarev and Obukhov~\cite{PO}, that the simplest metric--affine Lagrangian which depends on the tensor of homothetic curvature (this Lagrangian has the form of the Maxwell Lagrangian in which the electromagnetic field tensor is replaced by the tensor of homothetic curvature) generates the Einstein--Maxwell equations for the {\em massless} Weyl vector~\cite{HK,HLS} (corresponding to the photon).
Therefore the degrees of freedom contained in the affine connection provide an elegant way of unifying the classical gravitational and electromagnetic fields also in the metric--affine formulation of gravity.
The above Lagrangian was already mentioned, though without further investigation, by Hehl, Lord and Smalley~\cite{HLS} to remedy projective invariance of the gravitational action, and further studied by Vollick using a symmetric connection~\cite{Vol}.
A similar Lagrangian, with the electromagnetic field tensor represented by the curl of the torsion vector, was studied by Hammond~\cite{Ham}.
In Sec.~\ref{secSol} we present the general solution for the affine connection in the presence of matter fields that depend on the connection~\cite{PO}.

In Sec.~\ref{secMas} we use the Lagrange-multiplier method of Hehl and Kerlick~\cite{HK} to show that algebraic constraints on the torsion that break projective invariance of the connection and impose projective invariance on the tensor of homothetic curvature replace the massless Weyl vector with the {\em massive} Weyl vector related to the corresponding Lagrange multiplier.
The mass of this vector is given by the coupling of the affine electromagnetic Lagrangian to the gravitational field.
We examine several algebraic constraints on the torsion tensor.
We observe that the metric--affine formulation of gravity allows for a mechanism that {\em generates masses} of vectors, as it happens for electroweak gauge bosons via spontaneous symmetry breaking of the $SU(2)\times U(1)$ group~\cite{Wei}.
A geometrical mechanism of mass generation was already noted by Ponomarev and Obukhov~\cite{PO} who included the square of the torsion tensor in the gravitational Lagrangian, Hammond~\cite{Ham} who added the square of the torsion vector, and Vollick~\cite{Vol} and Hejna~\cite{Hej} who used a symmetric connection.
In this paper we treat this mechanism systematically using a Lagrange-multiplier formalism~\cite{HK} that allows us to study arbitrary constraints on the affine connection.
We briefly discuss and summarize the results in Sec.~\ref{secSum}.

\section{Field equations in metric--affine gravity}
\label{secField}

A general metric--affine Lagrangian density ${\cal L}$ depends on the affine connection $\Gamma^{\,\,\rho}_{\mu\,\nu}$ and the curvature tensor, $R^\rho_{\phantom{\rho}\mu\sigma\nu}=\Gamma^{\,\,\rho}_{\mu\,\nu,\sigma}-\Gamma^{\,\,\rho}_{\mu\,\sigma,\nu}+\Gamma^{\,\,\kappa}_{\mu\,\nu}\Gamma^{\,\,\rho}_{\kappa\,\sigma}-\Gamma^{\,\,\kappa}_{\mu\,\sigma}\Gamma^{\,\,\rho}_{\kappa\,\nu}$, as well as the symmetric metric tensor $g_{\mu\nu}$ of the Lorentzian signature $(+,-,-,-)$.
The antisymmetric part of the affine connection is the Cartan torsion tensor, $S^\rho_{\phantom{\rho}\mu\nu}=\Gamma^{\,\,\,\,\rho}_{[\mu\,\nu]}$, while its symmetric part can be split into the Christoffel--Levi-Civita connection, $\{^{\,\,\rho}_{\mu\,\nu}\}_g=\frac{1}{2}g^{\rho\lambda}(g_{\nu\lambda,\mu}+g_{\mu\lambda,\nu}-g_{\mu\nu,\lambda})$, and terms constructed from the torsion tensor and the nonmetricity tensor, $N_{\mu\nu\rho}=g_{\mu\nu;\rho}$~\cite{HK,HLS}:
\begin{equation}
\Gamma^{\,\,\rho}_{\mu\,\nu}=\{^{\,\,\rho}_{\mu\,\nu}\}_g+S^\rho_{\phantom{\rho}\mu\nu}+2S_{(\mu\nu)}^{\phantom{(\mu\nu)}\rho}+\frac{1}{2}N_{\mu\nu}^{\phantom{\mu\nu}\rho}-N^\rho_{\phantom{\rho}(\mu\nu)},
\label{split}
\end{equation}
where the semicolon denotes the covariant differentiation with respect to $\Gamma^{\,\,\rho}_{\mu\,\nu}$ (we use the notation of~\cite{Schr}).
The {\em Weyl vector} is defined as~\cite{HK,HLS}:
\begin{equation}
W_\nu=\frac{1}{2}(\Gamma^{\,\,\rho}_{\rho\,\nu}-\{^{\,\,\rho}_{\rho\,\nu}\}_g)=-\frac{1}{4}N^\rho_{\phantom{\rho}\rho\nu}.
\label{Weylvec1}
\end{equation}

We assume that the dependence of ${\cal L}$ on the curvature is restricted to the contracted curvature tensors~\cite{nonsym}, of which there exist three: the symmetric $P_{\mu\nu}=R_{(\mu\nu)}$ and antisymmetric $R_{[\mu\nu]}$ part of the Ricci tensor, $R_{\mu\nu}=R^\rho_{\phantom{\rho}\mu\rho\nu}$, and the antisymmetric {\em tensor of homothetic curvature}:
\begin{equation}
Q_{\mu\nu}=R^\rho_{\phantom{\rho}\rho\mu\nu}=\Gamma^{\,\,\rho}_{\rho\,\nu,\mu}-\Gamma^{\,\,\rho}_{\rho\,\mu,\nu},
\label{homoth}
\end{equation}
which has the form of a curl~\cite{Scho,Schr}.
The tensor of homothetic curvature is proportional to the curl of the Weyl vector:
\begin{equation}
Q_{\mu\nu}=2(W_{\nu,\mu}-W_{\mu,\nu}).
\label{Weylvec2}
\end{equation}
In addition, we assume that ${\cal L}$ does not depend on $R_{[\mu\nu]}$.
The simplest metric--affine Lagrangian density that depends on $P_{\mu\nu}$ is the Einstein--Palatini Lagrangian density of the gravitational field~\cite{FK3}:
\begin{equation}
{\cal L}_g=-\frac{1}{2\kappa}P_{\mu\nu}{\sf g}^{\mu\nu},
\label{LagGrav}
\end{equation}
where $\kappa=8\pi G$ ($c=1$), ${\sf g}^{\mu\nu}=\sqrt{-g}g^{\mu\nu}$ and $g=\mbox{det}(g_{\mu\nu})$.
The total Lagrangian density for the gravitational field and matter is given by ${\cal L}_g+{\cal L}_m$, where ${\cal L}_m$ is the Lagrangian density for matter that depends in general on both the metric and connection.
The variations of ${\cal L}_m$ with respect to the metric and connection define, respectively, the dynamical energy-momentum tensor $T_{\mu\nu}$:
\begin{equation}
T_{\mu\nu}=\frac{2}{\sqrt{-g}}\frac{\delta{\cal L}_m}{\delta g^{\mu\nu}},
\label{EMT}
\end{equation}
and the density conjugate to the connection (called hypermomentum in~\cite{HK,HLS}):
\begin{equation}
\Pi_{\phantom{\mu}\rho\phantom{\nu}}^{\mu\phantom{\rho}\nu}=-2\kappa\frac{\delta{\cal L}_m}{\delta \Gamma^{\,\,\rho}_{\mu\,\nu}},
\label{Pi}
\end{equation}
which has the same dimension as the connection.\footnote{
The variational derivative of a function ${\cal L}(\phi,\phi_{,\mu})$ with respect to a variable $\phi$ is defined as $\frac{\delta{\cal L}}{\delta\phi}=\frac{\partial{\cal L}}{\partial\phi}-(\frac{\partial{\cal L}}{\partial\phi_{,\mu}})_{,\mu}$.
}

The theory based on the Lagrangian density~(\ref{LagGrav}) without any constraints, i.e. with fully independent metric and connection, does not determine the connection uniquely because ${\cal L}_g$ is invariant under projective transformations of the connection~\cite{Hehl,HK,HLS,PO}:
\begin{equation}
\Gamma^{\,\,\rho}_{\mu\,\nu}\rightarrow\Gamma^{\,\,\rho}_{\mu\,\nu}+\delta^\rho_\mu\Lambda_\nu,
\label{proj}
\end{equation}
where $\Lambda_\nu$ is a vector function of the coordinates.
The same problem occurs if ${\cal L}_m$ does not depend on the connection, e.g., for electromagnetic fields or ideal fluids.
Therefore at least four degrees of freedom must be constrained to make such a theory consistent from a physical point of view~\cite{HLS}.
If, however, ${\cal L}_m$ does depend on the connection, e.g., for spinor fields or viscous fluids, the projective invariance of ${\cal L}_g$ imposes four algebraic constraints on the source density~(\ref{Pi}) and restricts forms of matter that can be described by the metric--affine formulation of gravity~\cite{HK,HLS,San}.
Consequently, we must change the form of the Lagrangian for the gravitational field so that it is not projectively invariant.

To break projective invariance of the field part of the Lagrangian we add a term that depends on the tensor of homothetic curvature $Q_{\mu\nu}$~\cite{HLS}.
The tensor $Q_{\mu\nu}$ is invariant only with respect to special projective transformations, i.e. projective transformations~(\ref{proj}) with $\Lambda_\nu=\lambda_{,\nu}$, where $\lambda$ is a scalar function of the coordinates~\cite{Kun}.
The simplest metric--affine Lagrangian density that depends on $Q_{\mu\nu}$, proposed by Hehl, Lord and Smalley~\cite{HLS}, Ponomarev and Obukhov~\cite{PO}, and Vollick~\cite{Vol}, has the form of the Maxwell Lagrangian for the electromagnetic field:
\begin{equation}
{\cal L}_Q=-\frac{\alpha}{4}\sqrt{-g}Q_{\mu\nu}Q^{\mu\nu},
\label{LagEM}
\end{equation}
where $\alpha$ is a constant.
Consequently, the total action is $S=\int d^4x({\cal L}_g+{\cal L}_Q+{\cal L}_m)$.
Varying $S$ with respect to the metric tensor and applying the principle of least action $\delta S=0$ for arbitrary variations of $g^{\mu\nu}$ yields the Einstein equations:
\begin{equation}
P_{\mu\nu}-\frac{1}{2}Pg_{\mu\nu}=\kappa\tilde{T}_{\mu\nu}=\kappa T_{\mu\nu}+\kappa\alpha\Bigl(\frac{1}{4}Q_{\rho\sigma}Q^{\rho\sigma}g_{\mu\nu}-Q_{\mu\rho}Q_\nu^{\phantom{\nu}\rho}\Bigr),
\label{Ein1}
\end{equation}
where $P=P_{\mu\nu}g^{\mu\nu}$ is the Ricci scalar.

The variation of $S$ with respect to the affine connection is $\delta S=-\frac{1}{2\kappa}\int d^4x({\sf g}^{\mu\nu}\delta P_{\mu\nu}+{\sf h}^{\mu\nu}\delta Q_{\mu\nu}+\Pi_{\phantom{\mu}\rho\phantom{\nu}}^{\mu\phantom{\rho}\nu}\delta\Gamma^{\,\,\rho}_{\mu\,\nu})$, where the antisymmetric tensor density ${\sf h}^{\mu\nu}$ is defined as~\cite{unif}:
\begin{equation}
{\sf h}^{\mu\nu}=-2\kappa\frac{\delta{\cal L}_Q}{\delta Q_{\mu\nu}}.
\label{as1}
\end{equation}
For the Lagrangian density~(\ref{LagEM}), the tensor density ${\sf h}^{\mu\nu}$ is linear in the tensor of homothetic curvature:
\begin{equation}
{\sf h}^{\mu\nu}=\kappa\alpha\sqrt{-g}Q^{\mu\nu}.
\label{as2}
\end{equation}
If we do not restrict the connection $\Gamma^{\,\,\rho}_{\mu\,\nu}$ to be symmetric, unlike in~\cite{Vol}, the variation of the Ricci tensor is given by the Palatini formula~\cite{Hehl,Scho,Schr}:
\begin{equation}
\delta R_{\mu\nu}=\delta\Gamma^{\,\,\rho}_{\mu\,\nu;\rho}-\delta\Gamma^{\,\,\rho}_{\mu\,\rho;\nu}-2S^\sigma_{\phantom{\sigma}\rho\nu}\delta\Gamma^{\,\,\rho}_{\mu\,\sigma}.
\label{Palat}
\end{equation}
Using the identity $\int d^4x({\sf V}^\mu)_{;\mu}=2\int d^4x S_\mu{\sf V}^\mu$~\cite{Schr}, where ${\sf V}^\mu$ is an arbitrary vector density and $S_\mu=S^\nu_{\phantom{\nu}\mu\nu}$ is the torsion vector, and applying the principle of least action $\delta S=0$ for arbitrary variations of $\Gamma^{\,\,\rho}_{\mu\,\nu}$, we obtain
\begin{equation}
{\sf g}^{\mu\nu}_{\phantom{\mu\nu};\rho}-{\sf g}^{\mu\sigma}_{\phantom{\mu\sigma};\sigma}\delta^\nu_\rho-2{\sf g}^{\mu\nu}S_\rho+2{\sf g}^{\mu\sigma}S_\sigma\delta^\nu_\rho+2{\sf g}^{\mu\sigma}S^\nu_{\phantom{\nu}\rho\sigma}=\Pi_{\phantom{\mu}\rho\phantom{\nu}}^{\mu\phantom{\rho}\nu}+2{\sf h}^{\nu\sigma}_{\phantom{\nu\sigma},\sigma}\delta^\mu_\rho.
\label{field1}
\end{equation}
This equation is equivalent to
\begin{equation}
{\sf g}^{\mu\nu}_{\phantom{\mu\nu},\rho}+\,^\ast\Gamma^{\,\,\mu}_{\sigma\,\rho}{\sf g}^{\sigma\nu}+\,^\ast\Gamma^{\,\,\nu}_{\rho\,\sigma}{\sf g}^{\mu\sigma}-\,^\ast\Gamma^{\,\,\sigma}_{\sigma\,\rho}{\sf g}^{\mu\nu}=\Pi_{\phantom{\mu}\rho\phantom{\nu}}^{\mu\phantom{\rho}\nu}-\frac{1}{3}\Pi_{\phantom{\mu}\sigma\phantom{\sigma}}^{\mu\phantom{\sigma}\sigma}\delta^\nu_\rho+2{\sf h}^{\nu\sigma}_{\phantom{\nu\sigma},\sigma}\delta^\mu_\rho-\frac{2}{3}{\sf h}^{\mu\sigma}_{\phantom{\mu\sigma},\sigma}\delta^\nu_\rho,
\label{field2}
\end{equation}
where $^\ast\Gamma^{\,\,\rho}_{\mu\,\nu}=\Gamma^{\,\,\rho}_{\mu\,\nu}+\frac{2}{3}\delta^\rho_\mu S_\nu$ is the projectively invariant connection (Schr\"{o}dinger's star-affinity)~\cite{Schr}.

Contracting the indices $\mu$ and $\rho$ in Eq.~(\ref{field1}) yields
\begin{equation}
{\sf h}^{\sigma\nu}_{\phantom{\sigma\nu},\sigma}={\sf j}^\nu,
\label{Max1}
\end{equation}
where
\begin{equation}
{\sf j}^\nu=\frac{1}{8}\Pi_{\phantom{\sigma}\sigma\phantom{\nu}}^{\sigma\phantom{\sigma}\nu}.
\label{Max2}
\end{equation}
Equation~(\ref{Max1}) has the form of the Maxwell equation for the electromagnetic field.
The density $\Pi_{\phantom{\mu}\rho\phantom{\nu}}^{\mu\phantom{\rho}\nu}$, in addition to $T_{\mu\nu}$, represents the {\em source} for the metric--affine field equations.
Since the tensor density ${\sf h}^{\mu\nu}$ is antisymmetric, the current vector density ${\sf j}^\mu$ must be conserved: ${\sf j}^\mu_{\phantom{\mu},\mu}=0$, which constrains how the connection $\Gamma^{\,\,\rho}_{\mu\,\nu}$ can enter a metric--affine Lagrangian density ${\cal L}$: $\Pi_{\phantom{\sigma}\sigma\phantom{\nu},\nu}^{\sigma\phantom{\sigma}\nu}=0$.
We note that if ${\cal L}$ does not depend on $Q_{\mu\nu}$, the field equation~(\ref{Max2}) becomes a stronger, algebraic constraint on how the Lagrangian depends on the connection: $\Pi_{\phantom{\sigma}\sigma\phantom{\nu}}^{\sigma\phantom{\sigma}\nu}=0$, which restricts forms of matter that can be described by the metric--affine formulation of gravity~\cite{HK,HLS,San}.
The dependence of a metric--affine Lagrangian on the tensor of homothetic curvature $Q_{\mu\nu}$, which will be associated with the electromagnetic field, {\em replaces} this unphysical constraint with a field equation for the tensor density ${\sf h}^{\mu\nu}$ (the Maxwell equations if this dependence is given by Eq.~(\ref{LagEM})).\footnote{
One way to overcome the constraint $\Pi_{\phantom{\sigma}\sigma\phantom{\nu}}^{\sigma\phantom{\sigma}\nu}=0$ is to restrict the torsion tensor to be traceless: $S_\mu=0$~\cite{San}.
This condition enters the Lagrangian density as a Lagrange multiplier term $\sqrt{-g}B^\mu S_\mu$, where the Lagrange multiplier $B^\mu$ is a vector.
Consequently, there is an extra term $-2\kappa\sqrt{-g}B^{[\mu}\delta^{\nu]}_\rho$ on the right-hand side of Eq.~(\ref{field1}) and Eq.~(\ref{Max2}) becomes ${\sf h}^{\sigma\nu}_{\phantom{\sigma\nu},\sigma}={\sf j}^\nu+\frac{3\kappa\sqrt{-g}}{8}B^\nu$.
Setting $B^\nu=-\frac{8}{3\kappa\sqrt{-g}}{\sf j}^\nu$ removes this constraint if ${\cal L}$ does not depend on $Q_{\mu\nu}$, or yields the wave equation ${\sf h}^{\sigma\nu}_{\phantom{\sigma\nu},\sigma}=0$ if ${\cal L}$ depends on $Q_{\mu\nu}$.
In both cases, the vector density ${\sf j}^\mu$ does not need to be conserved.
Therefore introducing the dependence of ${\cal L}$ on $Q_{\mu\nu}$ is more suitable than imposing $S_\mu=0$ if we want to unify gravitation and electromagnetism in the metric--affine formalism.
Another way to overcome the constraint $\Pi_{\phantom{\sigma}\sigma\phantom{\nu}}^{\sigma\phantom{\sigma}\nu}=0$ is to restrict the Weyl vector to vanish: $W_\mu=0$~\cite{HLS}.
\label{foot}
}

The tensor $P_{\mu\nu}$ is invariant under an infinitesimal projective transformation~(\ref{proj}): $\delta\Gamma^{\,\,\rho}_{\mu\,\nu}=\delta^\rho_\mu\delta V_\nu$.\footnote{
Equation~(\ref{Palat}) yields $\delta P_{\mu\nu}=0$ for $\delta\Gamma^{\,\,\rho}_{\mu\,\nu}=\delta^\rho_\mu\delta V_\nu$.
}
Under the same transformation, the tensor $Q_{\mu\nu}$ changes according to $Q_{\mu\nu}\rightarrow Q_{\mu\nu}+4(\delta V_{\nu,\mu}-\delta V_{\mu,\nu})$.
Consequently, the action changes according to $\delta S=-\frac{1}{2\kappa}\int d^4x(\Pi_{\phantom{\mu}\rho\phantom{\nu}}^{\mu\phantom{\rho}\nu}\delta\Gamma^{\,\,\rho}_{\mu\,\nu}+{\sf h}^{\mu\nu}\delta Q_{\mu\nu})=-\frac{1}{2\kappa}\int d^4x(\Pi_{\phantom{\sigma}\sigma\phantom{\mu}}^{\sigma\phantom{\sigma}\mu}+8{\sf h}^{\mu\nu}_{\phantom{\mu\nu},\nu})\delta V_\mu$.
This expression is identically zero due to the field equation~(\ref{Max1}) so, although ${\cal L}_m$ and ${\cal L}_Q$ are not projectively invariant, the total action is.
The formal similarity between the tensor of homothetic curvature $Q_{\mu\nu}$ and the electromagnetic field tensor $F_{\mu\nu}=A_{\nu,\mu}-A_{\mu,\nu}$ (both tensors are curls) suggests that they are proportional to one another~\cite{Goe,FK2}.
If $\hbar=1$ then the constant of proportionality has the dimension of electric charge~\cite{FK2} and, without loss of generality, can be taken equal to the charge of the electron $e$:
\begin{equation}
F_{\mu\nu}=eQ_{\mu\nu}.
\label{prop}
\end{equation}
Accordingly, the Weyl vector is proportional to the electromagnetic four-potential $A_\mu$:
\begin{equation}
W_\mu=\frac{1}{2e}A_\mu.
\label{prop2}
\end{equation}

The degrees of freedom contained in the affine connection provide an elegant way of unifying the classical gravitational and electromagnetic fields in the metric--affine formulation of gravity.
In addition, the Maxwell equations constrain the desired four degrees of freedom so that the metric--affine formulation of gravity is consistent from a physical point of view~\cite{HLS}.
Finally, we can interpret the electromagnetic field in a metric--affine gravity as a field whose role is to {\em restore} projective invariance of a metric--affine Lagrangian broken by matter terms that depend explicitly on the affine connection, without constraining {\em ad hoc} the connection~\cite{unif}.

\section{Solution of field equations}
\label{secSol}

Substituting Eq.~(\ref{Max2}) to~(\ref{field2}) and symmetrizing the indices $\mu$ and $\nu$ yield
\begin{equation}
{\sf g}^{\mu\nu}_{\phantom{\mu\nu},\rho}+\,^\ast\Gamma^{\,\,\,\,\mu}_{(\sigma\,\rho)}{\sf g}^{\sigma\nu}+\,^\ast\Gamma^{\,\,\,\,\nu}_{(\rho\,\sigma)}{\sf g}^{\mu\sigma}-\,^\ast\Gamma^{\,\,\,\,\sigma}_{(\sigma\,\rho)}{\sf g}^{\mu\nu}=\Sigma_{\phantom{\mu}\rho\phantom{\nu}}^{\mu\phantom{\rho}\nu},
\label{field3}
\end{equation}
where
\begin{equation}
\Sigma_{\phantom{\mu}\rho\phantom{\nu}}^{\mu\phantom{\rho}\nu}=\Pi_{\phantom{(\mu}\rho\phantom{\nu)}}^{(\mu\phantom{\rho}\nu)}-\frac{1}{3}\delta^{(\mu}_\rho\Pi_{\phantom{\nu)}\sigma\phantom{\sigma}}^{\nu)\phantom{\sigma}\sigma}-\frac{1}{6}\Pi_{\phantom{\sigma}\sigma\phantom{(\mu}}^{\sigma\phantom{\sigma}(\mu}\delta^{\nu)}_\rho.
\label{sigma}
\end{equation}
Equation~(\ref{field3}) is algebraic and linear in $^\ast\Gamma^{\,\,\,\,\rho}_{(\mu\,\nu)}$ as a function of the metric tensor, its first derivatives and the density $\Pi_{\phantom{\mu}\rho\phantom{\nu}}^{\mu\phantom{\rho}\nu}$.
We decompose the connection $^\ast\Gamma^{\,\,\rho}_{\mu\,\nu}$ as~\cite{unif}:
\begin{equation}
^\ast\Gamma^{\,\,\rho}_{\mu\,\nu}=\{^{\,\,\rho}_{\mu\,\nu}\}_g+V^\rho_{\phantom{\rho}\mu\nu},
\label{sol1}
\end{equation}
where $V^\rho_{\phantom{\rho}\mu\nu}=S^\rho_{\phantom{\rho}\mu\nu}+2S_{(\mu\nu)}^{\phantom{(\mu\nu)}\rho}+\frac{1}{2}N_{\mu\nu}^{\phantom{\mu\nu}\rho}-N^\rho_{\phantom{\rho}(\mu\nu)}+\frac{2}{3}\delta^\rho_\mu S_\nu$ is a projectively invariant deflection tensor.
Consequently, the Ricci tensor $R_{\mu\nu}$ associated with the affine connection $\Gamma^{\,\,\rho}_{\mu\,\nu}$ is given by~\cite{Scho}:
\begin{equation}
R_{\mu\nu}=R_{\mu\nu}^{(g)}-\frac{2}{3}(S_{\nu:\mu}-S_{\mu:\nu})+V^\rho_{\phantom{\rho}\mu\nu:\rho}-V^\rho_{\phantom{\rho}\mu\rho:\nu}+V^\sigma_{\phantom{\sigma}\mu\nu}V^\rho_{\phantom{\rho}\sigma\rho}-V^\sigma_{\phantom{\sigma}\mu\rho}V^\rho_{\phantom{\rho}\sigma\nu},
\label{sol2}
\end{equation}
where $R_{\mu\nu}^{(g)}$ is the Riemannian Ricci tensor constructed from the metric tensor $g_{\mu\nu}$ and the colon denotes the covariant differentiation with respect to $\{^{\,\,\rho}_{\mu\,\nu}\}_g$.
Equation~(\ref{Ein1}) and symmetrized Eq.~(\ref{sol2}) give the Einstein equations in the purely metric formulation:
\begin{equation}
R_{\mu\nu}^{(g)}-\frac{1}{2}R^{(g)}g_{\mu\nu}=\kappa(\tilde{T}_{\mu\nu}+\Theta_{\mu\nu}),
\label{Ein2}
\end{equation}
where $\Theta_{\mu\nu}$ is the effective energy-momentum tensor generated by the affine connection:
\begin{equation}
\Theta_{\mu\nu}=\frac{1}{\kappa}(-V^\rho_{\phantom{\rho}(\mu\nu):\rho}+V^\rho_{\phantom{\rho}(\mu|\rho|:\nu)}-V^\sigma_{\phantom{\sigma}(\mu\nu)}V^\rho_{\phantom{\rho}\sigma\rho}+V^\sigma_{\phantom{\sigma}(\mu|\rho}V^\rho_{\phantom{\rho}\sigma|\nu)})-\frac{1}{2\kappa}g_{\mu\nu}(2V^{\rho\sigma}_{\phantom{\rho\sigma}[\rho:\sigma]}-V^{\sigma\tau}_{\phantom{\sigma\tau}\tau}V^\rho_{\phantom{\rho}\sigma\rho}+V^{\sigma\tau}_{\phantom{\sigma\tau}\rho}V^\rho_{\phantom{\rho}\sigma\tau}).
\label{Ein3}
\end{equation}
The Bianchi identity $(R^{\mu\nu(g)}-\frac{1}{2}R^{(g)}g^{\mu\nu})_{:\nu}=0$ yields the covariant conservation of the total energy--momentum tensor: $(\tilde{T}^{\mu\nu}+\Theta^{\mu\nu})_{:\nu}=0$.

Substituting Eq.~(\ref{sol1}) to~(\ref{field3}) gives
\begin{equation}
V^\mu_{\phantom{\mu}(\sigma\rho)}{\sf g}^{\sigma\nu}+V^\nu_{\phantom{\nu}(\rho\sigma)}{\sf g}^{\mu\sigma}-V^\sigma_{\phantom{\sigma}(\sigma\rho)}{\sf g}^{\mu\nu}=\Sigma_{\phantom{\mu}\rho\phantom{\nu}}^{\mu\phantom{\rho}\nu}.
\label{sol3}
\end{equation}
Its solution is:
\begin{equation}
V^\rho_{\phantom{\rho}(\mu\nu)}=\frac{1}{2\sqrt{-g}}(\Delta_{\phantom{\rho}\nu\phantom{\sigma}}^{\rho\phantom{\nu}\sigma}g_{\mu\sigma}+\Delta_{\phantom{\rho}\mu\phantom{\sigma}}^{\rho\phantom{\mu}\sigma}g_{\nu\sigma}-\Delta_{\phantom{\alpha}\gamma\phantom{\beta}}^{\alpha\phantom{\gamma}\beta}g_{\mu\alpha}g_{\nu\beta}g^{\rho\gamma}),
\label{sol4}
\end{equation}
where
\begin{equation}
\Delta_{\phantom{\mu}\rho\phantom{\nu}}^{\mu\phantom{\rho}\nu}=\Sigma_{\phantom{\mu}\rho\phantom{\nu}}^{\mu\phantom{\rho}\nu}-\frac{1}{2}\Sigma_{\phantom{\alpha}\rho\phantom{\beta}}^{\alpha\phantom{\rho}\beta}g_{\alpha\beta}g^{\mu\nu}.
\label{delta}
\end{equation}
Substituting Eq.~(\ref{Max2}) to~(\ref{field2}) and antisymmetrizing the indices $\mu$ and $\nu$ yield
\begin{equation}
V^\mu_{\phantom{\mu}[\sigma\rho]}{\sf g}^{\nu\sigma}-V^\nu_{\phantom{\nu}[\sigma\rho]}{\sf g}^{\mu\sigma}=\Omega_\rho^{\phantom{\rho}\mu\nu},
\label{field4}
\end{equation}
where
\begin{equation}
\Omega_\rho^{\phantom{\rho}\mu\nu}=\Pi_{\phantom{[\mu}\rho\phantom{\nu]}}^{[\mu\phantom{\rho}\nu]}-\frac{1}{3}\Pi_{\phantom{[\sigma}\sigma\phantom{\nu]}}^{[\sigma\phantom{\sigma}\nu]}\delta^\mu_\rho+\frac{1}{3}\Pi_{\phantom{[\sigma}\sigma\phantom{\mu]}}^{[\sigma\phantom{\sigma}\mu]}\delta^\nu_\rho
\label{omega}
\end{equation}
is a traceless tensor density.
Consequently, we find
\begin{equation}
V^\rho_{\phantom{\rho}[\mu\nu]}=\frac{1}{2\sqrt{-g}}(\Omega_\nu^{\phantom{\nu}\rho\sigma}g_{\mu\sigma}-\Omega_\mu^{\phantom{\mu}\rho\sigma}g_{\nu\sigma}-\Omega_\gamma^{\phantom{\gamma}\alpha\beta}g_{\mu\alpha}g_{\nu\beta}g^{\rho\gamma}).
\label{sol5}
\end{equation}
Equations~(\ref{sol4}) and~(\ref{sol5}) give the tensor $V^\rho_{\phantom{\rho}\mu\nu}$ which is linear in $\Pi_{\phantom{\mu}\rho\phantom{\nu}}^{\mu\phantom{\rho}\nu}$~\cite{unif}.
A general solution for the affine connection in the presence of matter fields that depend on the connection is also given in~\cite{PO}.

Equation~(\ref{sol1}) gives the following expression for the tensor of homothetic curvature~\cite{Scho}:
\begin{equation}
Q_{\mu\nu}=-\frac{8}{3}(S_{\nu,\mu}-S_{\mu,\nu})+V^\rho_{\phantom{\rho}\rho\nu,\mu}-V^\rho_{\phantom{\rho}\rho\mu,\nu}.
\label{q1}
\end{equation}
Accordingly, the Weyl vector is:
\begin{equation}
W_\mu=-\frac{4}{3}S_\mu+\frac{1}{2}V^\rho_{\phantom{\rho}\rho\mu}.
\label{Wey}
\end{equation}
The torsion vector is related to the electromagnetic potential $A_\mu$ via Eq.~(\ref{q1}) and the correspondence relation~(\ref{prop}):\footnote{
The torsion vector in~\cite{Ham} is proportional to the electromagnetic potential.
}
\begin{equation}
S_\nu=\frac{3}{8}\Bigl(-\frac{A_\nu}{e}+V^\rho_{\phantom{\rho}\rho\nu}\Bigr).
\label{tor}
\end{equation}
Gauge transformations $A_\nu\rightarrow A_\nu+\lambda_{,\nu}$ (equivalent to special projective transformations) do not affect Eq.~(\ref{q1}) and the total action.
Equations~(\ref{as2}) and~(\ref{Max1}) reproduce the Maxwell equation with source, $\frac{1}{\sqrt{-g}}(\sqrt{-g}F_{\alpha\beta}g^{\nu\alpha}g^{\mu\beta})_{,\nu}=j^\mu$, where $j^\mu$ is the electromagnetic current four-vector~\cite{LL2}, if we identify:
\begin{equation}
{\sf j}^\mu=\frac{\kappa\alpha}{e}\sqrt{-g}j^\mu.
\label{iden}
\end{equation}

If there are no sources, $\Pi_{\phantom{\mu}\rho\phantom{\nu}}^{\mu\phantom{\rho}\nu}=0$, the connection $\Gamma^{\,\,\rho}_{\mu\,\nu}$ depends only on the metric tensor $g_{\mu\nu}$ representing the gravitational field and the torsion vector $S_\mu$ proportional to the electromagnetic potential~\cite{FK2,Ham} and thus to the Weyl vector.
If we impose the Lorentz condition $W^\mu_{\phantom{\mu}:\mu}=0$ (by applying a suitable special projective transformation), Eqs.~(\ref{as2}),~(\ref{Max1}) and~(\ref{q1}) yield the wave equation for the massless Weyl vector:
\begin{equation}
\triangle_g W^\mu-R^\mu_{\phantom{\mu}\nu(g)}W^\nu=0,
\label{q2}
\end{equation}
where $\triangle_g$ denotes the covariant d'Alembert operator with respect to $\{^{\,\,\rho}_{\mu\,\nu}\}_g$.

Lastly, we consider the case when the source density~(\ref{Pi}) is antisymmetric (which we study in the next section):
\begin{equation}
\Pi_{\phantom{\mu}\rho\phantom{\nu}}^{\mu\phantom{\rho}\nu}=\Pi_{\phantom{[\mu}\rho\phantom{\nu]}}^{[\mu\phantom{\rho}\nu]}.
\label{Pi2}
\end{equation}
Equations~(\ref{sigma}),~(\ref{sol4}) and~(\ref{delta}) give $V^\rho_{\phantom{\rho}(\rho\nu)}=-\frac{2}{3\sqrt{-g}}g_{\mu\nu}{\sf j}^\mu$, while Eqs.~(\ref{omega}) and~(\ref{sol5}) give $V^\rho_{\phantom{\rho}[\rho\nu]}=0$.
Therefore we find
\begin{equation}
V^\rho_{\phantom{\rho}\rho\nu}=-\frac{2\kappa\alpha}{3e}j_\nu,
\label{sol6}
\end{equation}
and the tensor of homothetic curvature is:
\begin{equation}
Q_{\mu\nu}=-\frac{8}{3}(S_{\nu,\mu}-S_{\mu,\nu})-\frac{2\kappa\alpha}{3e}(j_{\nu,\mu}-j_{\mu,\nu}).
\label{q3}
\end{equation}
The torsion vector turns out to be, due to Eq.~(\ref{tor}), a linear combination of the electromagnetic potential and the electromagnetic current:
\begin{equation}
S_\nu=-\frac{3}{8e}A_\nu-\frac{\kappa\alpha}{4e}j_\nu.
\label{q4}
\end{equation}
The Weyl vector, proportional to the electromagnetic potential due to Eq.~(\ref{prop2}), satisfies now
\begin{equation}
\triangle_g W^\mu-R^\mu_{\phantom{\mu}\nu(g)}W^\nu=\frac{j^\mu}{2e},
\label{q5}
\end{equation}
playing the role of the {\em massless} vector (photon) generated by the tensor of homothetic curvature in a metric--affine Lagrangian in the presence of electromagnetic sources.

\section{Generation of mass from algebraic constraints on torsion}
\label{secMas}

Combining Eqs.~(\ref{as2}),~(\ref{Max1}),~(\ref{iden}) and~(\ref{q3}) gives
\begin{equation}
-\Bigl(\frac{8e}{3}(S^{\nu,\mu}-S^{\mu,\nu})+\frac{2\kappa\alpha}{3}(j^{\nu,\mu}-j^{\mu,\nu})\Bigr)_{:\mu}=j^\nu.
\label{Max3}
\end{equation}
This equation formally looks like a dynamical equation for the vector field $j^\mu$.
However, the vector $j^\mu$ is not a dynamical field but constitutes the source for the dynamical field $W_\mu$ (or $A_\mu$) since Eq.~(\ref{Max3}) reads
\begin{equation}
2e(W^{\nu,\mu}-W^{\mu,\nu})_{:\mu}=j^\nu,
\label{Max4}
\end{equation}
or Eq.~(\ref{q5}).
We can ask whether it is possible to turn the vector $j^\mu$ into a dynamical field and thus modify Eq.~(\ref{Max4}).
The answer is yes and the mechanism to dynamize the source $j^\mu$ is related to constraining the torsion tensor.
To constrain the torsion we use the Lagrange-multiplier method of Hehl and Kerlick~\cite{HK} that allows to study arbitrary constraints on the affine connection.

\subsection{Torsionless connection}
\label{TC}

Let us consider the strongest constraint on the torsion tensor:
\begin{equation}
S^\rho_{\phantom{\rho}\mu\nu}=0.
\label{cond1}
\end{equation}
This condition enters the Lagrangian density as the Lagrange-multiplier term, $\sqrt{-g}\beta_\rho^{\phantom{\rho}\mu\nu}S^\rho_{\phantom{\rho}\mu\nu}$, where the Lagrange multiplier $\beta_\rho^{\phantom{\rho}\mu\nu}$ is a tensor field antisymmetric in the indices $\mu$ and $\nu$:
\begin{equation}
{\cal L}=-\frac{1}{2\kappa}P_{\mu\nu}{\sf g}^{\mu\nu}-\frac{\alpha}{4}\sqrt{-g}Q_{\mu\nu}Q^{\mu\nu}+\sqrt{-g}\beta_\rho^{\phantom{\rho}\mu\nu}S^\rho_{\phantom{\rho}\mu\nu}+{\cal L}_m.
\label{mas1}
\end{equation}
Varying the corresponding action with respect to the field $\beta_\rho^{\phantom{\rho}\mu\nu}$ gives Eq.~(\ref{cond1}) and thus $S_\mu=0$.
Let us also assume, for simplicity, the absence of matter fields, ${\cal L}_m=0$.
To solve the field equation related to the variation of the connection, we treat the Lagrange-multiplier term as the effective Lagrangian density for matter.\footnote{
In the following text, $j^\nu$ denotes not the electromagnetic current (which is zero in the absence of matter fields) but the effective current four-vector associated with a Lagrange-multiplier constraint on the connection.
}
Consequently, we find
\begin{equation}
\Pi_{\phantom{\mu}\rho\phantom{\nu}}^{\mu\phantom{\rho}\nu}=-2\kappa\sqrt{-g}\beta_\rho^{\phantom{\rho}\mu\nu}
\label{mas2}
\end{equation}
and, because of Eq.~(\ref{iden}),
\begin{equation}
j^\nu=-\frac{2e}{\alpha}\beta_\rho^{\phantom{\rho}\rho\nu}.
\label{mas3}
\end{equation}
Accordingly, the Weyl vector is given by Eqs.~(\ref{Wey}) and~(\ref{sol6}):
\begin{equation}
W_\nu=-\frac{\kappa\alpha}{3e}j_\nu.
\label{sol7}
\end{equation}

The source density~(\ref{mas2}) is antisymmetric so we can apply the results of the preceding section and Eq.~({\ref{Max4}), obtaining
\begin{equation}
(W^{\nu,\mu}-W^{\mu,\nu})_{:\mu}=-\frac{3}{2\kappa\alpha}W^\nu.
\label{Max5}
\end{equation}
The Lorentz condition $W^\mu_{\phantom{\mu}:\mu}=0$ is satisfied automatically, yielding the Einstein--Proca equation for the {\em massive} Weyl vector:\footnote{
The massive Weyl vector loses the meaning of the quantity representing the electromagnetic potential.
}
\begin{equation}
\triangle_g W^\mu-R^\mu_{\phantom{\mu}\nu(g)}W^\nu+m_W^2W^\mu=0,
\label{Proc}
\end{equation}
with mass
\begin{equation}
m_W=\sqrt{\frac{3}{2\kappa\alpha}}.
\label{mas}
\end{equation}
The mass of the vector $W^\mu$ is determined by the coupling constant $\alpha$.
The same result was observed by Vollick who used the symmetric connection in a variational principle~\cite{Vol}.

\subsection{Traceless torsion}
\label{TT}

We now consider a weaker constraint on the torsion tensor, imposing its trace to vanish:
\begin{equation}
S_\mu=0.
\label{cond2}
\end{equation}
This condition enters the Lagrangian density as the Lagrange-multiplier term: $\sqrt{-g}B^\mu S_\mu$ (cf. footnote \ref{foot}), so the total Lagrangian density is:
\begin{equation}
{\cal L}=-\frac{1}{2\kappa}P_{\mu\nu}{\sf g}^{\mu\nu}-\frac{\alpha}{4}\sqrt{-g}Q_{\mu\nu}Q^{\mu\nu}+\sqrt{-g}B^\mu S_\mu,
\label{mas4}
\end{equation}
since varying the corresponding action with respect to the field $B^\mu$ gives Eq.~(\ref{cond2}).
We again treat the Lagrange-multiplier term as the matter part of the Lagrangian, obtaining
\begin{equation}
\Pi_{\phantom{\mu}\rho\phantom{\nu}}^{\mu\phantom{\rho}\nu}=-2\kappa\sqrt{-g}B^{[\mu}\delta^{\nu]}_\rho
\label{mas5}
\end{equation}
and
\begin{equation}
j^\nu=\frac{3e}{8\alpha}B^\nu.
\label{mas6}
\end{equation}
The source density~(\ref{mas5}) is again antisymmetric so we can apply the results of the preceding section and Eqs.~(\ref{Max4}) and~({\ref{sol7}), obtaining the Einstein--Proca equations~(\ref{Max5}) and~({\ref{Proc}) for the {\em massive} Weyl vector of mass~(\ref{mas}).

We obtain the same results if the torsion vector is a gradient of a scalar:
\begin{equation}
S_\mu=\lambda_{,\mu},
\label{cond4}
\end{equation}
which is weaker than the constraint~(\ref{cond2}).
Therefore it is the constraint $S_\mu=\lambda_{,\mu}$, or equivalently $S_{\nu,\mu}-S_{\mu,\nu}=0$, that gives mass to the Weyl vector in our Lagrangian.
From Eq.~(\ref{q1}) it follows that this condition causes the tensor of homothetic curvature to depend only on the projectively invariant tensor $V^\rho_{\phantom{\rho}\mu\nu}$, i.e. {\em imposing projective invariance on the tensor of homothetic curvature gives mass to the vector degree of freedom} of our Lagrangian.

\subsection{Vector torsion}
\label{VT}

We now consider a different constraint on the torsion tensor:
\begin{equation}
S^\rho_{\phantom{\rho}\mu\nu}=\frac{2}{3}S_{[\mu}\delta_{\nu]}^\rho,
\label{cond3}
\end{equation}
i.e. the torsion tensor depends completely on the torsion vector.\footnote{
Contracting the indices $\mu$ and $\rho$ yields the identity.
}
This condition enters the Lagrangian density as the Lagrange-multiplier term: $\sqrt{-g}\gamma_\rho^{\phantom{\rho}\mu\nu}(S^\rho_{\phantom{\rho}\mu\nu}-\frac{2}{3}S_{[\mu}\delta_{\nu]}^\rho)$, where the Lagrange multiplier $\gamma_\rho^{\phantom{\rho}\mu\nu}$ is a tensor field antisymmetric in the indices $\mu$ and $\nu$:
\begin{equation}
{\cal L}=-\frac{1}{2\kappa}P_{\mu\nu}{\sf g}^{\mu\nu}-\frac{\alpha}{4}\sqrt{-g}Q_{\mu\nu}Q^{\mu\nu}+\sqrt{-g}\gamma_\rho^{\phantom{\rho}\mu\nu}\Bigl(S^\rho_{\phantom{\rho}\mu\nu}-\frac{2}{3}S_{[\mu}\delta_{\nu]}^\rho\Bigr).
\label{mas7}
\end{equation}
Varying the corresponding action with respect to the field $\gamma_\rho^{\phantom{\rho}\mu\nu}$ reproduces Eq.~(\ref{cond3}).
We again treat the Lagrange-multiplier term as the matter part of the Lagrangian, obtaining
\begin{equation}
\Pi_{\phantom{\mu}\rho\phantom{\nu}}^{\mu\phantom{\rho}\nu}=-2\kappa\sqrt{-g}\Bigl(\gamma_\rho^{\phantom{\rho}\mu\nu}+\frac{2}{3}\gamma_\sigma^{\phantom{\sigma}\sigma[\mu}\delta^{\nu]}_\rho\Bigr).
\label{mas8}
\end{equation}
The effective current vanishes identically:
\begin{equation}
j^\nu=0.
\label{mas9}
\end{equation}
Accordingly, the Weyl vector is given by Eqs.~(\ref{Wey}) and~(\ref{sol6}):
\begin{equation}
W_\nu=-\frac{4}{3}S_\nu,
\label{sol8}
\end{equation}
and is {\em massless}, satisfying
\begin{equation}
\triangle_g W^\mu-R^\mu_{\phantom{\mu}\nu(g)}W^\nu=0.
\label{q6}
\end{equation}

The condition~(\ref{cond3}) is the only algebraic constraint on the connection that is projectively invariant and contains only the torsion tensor.
Therefore we can state that {\em constraints that break projective invariance generate mass of the vector} associated with the dynamical connection (the Weyl vector), while {\em projectively invariant algebraic constraints on the connection do not generate mass} of this vector.
From Eq.~(\ref{q1}) it follows that this condition causes the tensor of homothetic curvature to depend on the torsion vector which is not projectively invariant. 
Since we do not impose projective invariance on the tensor of homothetic curvature, the vector degree of freedom remains massless (cf. Subsec.~\ref{TT}).

\subsection{Nonlinear constraints on torsion}
\label{ST}

We examined three simplest (linear) constraints on the the torsion.
Let us consider the simplest quadratic constraint:
\begin{equation}
S^\mu S_\mu=0,
\label{cond5}
\end{equation}
introducing it through the Lagrange-multiplier term, $\sqrt{-g}C S^\mu S_\mu$.
Varying the corresponding action with respect to the field $C$ reproduces Eq.~(\ref{cond5}).
Following the previous subsections, we find
\begin{equation}
\Pi_{\phantom{\mu}\rho\phantom{\nu}}^{\mu\phantom{\rho}\nu}=-4\kappa\sqrt{-g}CS^{[\mu}\delta^{\nu]}_\rho
\label{qu1}
\end{equation}
and
\begin{equation}
j^\nu=\frac{3e}{4\alpha}CS^\nu.
\label{qu2}
\end{equation}
Accordingly, the Weyl vector is given by
\begin{equation}
W_\nu=-\Bigl(\frac{4}{3}+\frac{\kappa C}{4}\Bigr)S_\nu
\label{qu3}
\end{equation}
and satisfies
\begin{equation}
W^\mu W_\mu=0.
\label{nul}
\end{equation}
Equation~(\ref{q5}) reads
\begin{equation}
\triangle_g W^\mu-R^\mu_{\phantom{\mu}\nu(g)}W^\nu=-\frac{9C}{32\alpha(1+\frac{3\kappa C}{16})}W^\mu.
\label{q7}
\end{equation}

Equation~(\ref{q7}) has the form of the Einstein--Proca equation~(\ref{Proc}) with an effective mass
\begin{equation}
m_W=\sqrt{\frac{9C}{32\alpha(1+\frac{3\kappa C}{16})}},
\label{run}
\end{equation}
which depends on the scalar field $C$.
The field $C$ replaces the degree of freedom that is constrained by Eq.~(\ref{nul}).
The effective mass of the Weyl vector ranges from $m_W=0$ for $C=0$ to $m_W=\sqrt{\frac{3}{2\kappa\alpha}}$, as in Eq.~(\ref{mas}), for $C\rightarrow\infty$.
The dependence of the effective mass on the scalar degree of freedom is a consequence of the non-zero torsion vector $S_\mu$.
Nonlinear constraints, unlike those in Eqs.~(\ref{cond1}) and~(\ref{cond2}), do not impose vanishing of this vector so the Weyl vector is a linear combination of $S_\mu$ and the effective current.
If a nonlinear constraint on the torsion has the form of a scalar, the corresponding Lagrange multiplier is a scalar density and the resulting degree of freedom that appears in place of this constraint is a scalar which enters the effective current and effective mass.
Since the field equations arising from nonlinear constraints on the torsion are not simple Einstein--Maxwell or Einstein--Proca equations, we conclude that these constraints do not correspond to physical conditions (vector particles with constant masses).
Similar conclusions can be made for nonlinear constraints on the connection that have the form of tensors and for constraints that contain covariant derivatives.

\section{Discussion and summary}
\label{secSum}

The degrees of freedom contained in the affine connection not only provide an elegant way of unifying the classical gravitational and electromagnetic fields within the metric--affine formulation of gravity, but also can incorporate massive vectors.
The simplest metric--affine Lagrangian that depends on the tensor of homothetic curvature generates the Einstein--Maxwell equations for the electromagnetic potential corresponding to the massless photon.
In the previous section we examined three linear constraints on the antisymmetric part of the affine connection (the torsion tensor).
We found that constraints that break projective invariance generate the Einstein--Proca equations for a massive vector which appears in place of the photon.
Therefore projective-invariance breaking generates masses of vectors in metric--affine gravity that incorporates the tensor of homothetic curvature.
We also examined one simple quadratic constraint on the torsion and concluded that nonlinear constraints generate the Einstein--Proca equations with running masses that do not correspond to physical vector particles.

Although we only considered algebraic constraints on the torsion, the Lagrange-multiplier formalism allows to study arbitrary constraints on the affine connection.
We can, e.g., require the connection to be metric-compatible~\cite{HK}: $g_{\mu\nu;\rho}=0$, which implies $\Gamma^{\,\,\rho}_{\rho\,\nu}=\{^{\,\,\rho}_{\rho\,\nu}\}_g$ and thus $Q_{\mu\nu}=0$.
Therefore, under this condition, the electromagnetic field loses its association with the affine connection since imposing the metric-compatibility on the connection means that we are dealing with the purely metric formulation of gravity rather than metric--affine.
We can also introduce differential constraints on the torsion, such as $S_{\mu;\nu}=0$.
These constraints, however, impose {\em ad hoc} differential equations on the connection which should result from a variational principle with respect to a geometrical quantity rather than from Lagrange multipliers.
Therefore the only physical constraints on the connection in the metric--affine formulation of gravity should have the form of linear algebraic equations for the torsion tensor.

The presented paper is a mathematical exercise that can give insights into possible geometrical integration of the bosonic part of the Glashow--Weinberg--Salam electroweak model~\cite{Wei,EW} with gravitation.
In the electroweak model, masses of gauge bosons arise naturally from spontaneous symmetry breaking (the Higgs mechanism) of the $SU(2)\times U(1)$ group~\cite{Wei,Hig}.
Gauge-boson masses have also been shown to arise when gauge invariance is spontaneously broken in higher-dimensional theories of gravity~\cite{GEW}.
In this paper we showed that projective-symmetry breaking can generate massive gauge bosons in four dimensions.
To include four gauge bosons instead of one in the metric--affine formulation of gravity (and generate masses for three of them) we should introduce the covariant derivative acting on the electroweak spinor doublet (the fermionic part)~\cite{Wei}.
Since the metric tensor and connection cannot describe spinors in curved spacetime, we should use the Einstein--Cartan--Kibble--Sciama formulation of gravity with the tetrad and spin connection as dynamical variables~\cite{KS} and impose constraints on the Fock--Ivanenko coefficients that appear in the covariant derivative of a spinor~\cite{Lord}.

\end{document}